\documentclass[manuscript]{aastex} 
 \slugcomment{Not
to appear in Nonlearned J., 45.}  \shorttitle{Collapsed Cores in
Globular Clusters} \shortauthors{Djorgovski et al.}

\begin{document}

\title{NUCLEAR ACTIVITY IN ISOLATED GALAXIES}

\author{Francisco J. Hern\'andez-Ibarra\altaffilmark{1}, Deborah
Dultzin\altaffilmark{1}, Yair Krongold\altaffilmark{1}}
\affil{$^{1}$Instituto de Astronom\'ia, Universidad Nacional
Aut\'onoma de M\'exico, Apartado Postal 70-264, 04510 M\'exico DF,
M\'exico.}  \author{Ascensi\'on Del Olmo\altaffilmark{2}, Jaime
Perea\altaffilmark{2} \& J. J. Gonz\'alez\altaffilmark{1}}
\affil{$^{2}$Instituto de Astrof\'isica de Andaluc\'ia (C.S.I.C.)
Apartado 3004, 18080 Granada, Spain} \email{hibarra@astro.unam.mx}

\begin{abstract}

We present a spectroscopic study of the incidence of AGN nuclear
activity in two samples of isolated galaxies (Karachentseva, V.E. \&
Varela, J.).  Our results show that the incidence of non-thermal
nuclear activity is about 43\% and 31\% for galaxies with emission
lines and for the total sample 40\% and 27\% respectively. For the
first time we have a large number of bona-fide isolated galaxies (513
objects), with statistically significant number of all morphological
types. A large fraction ($\sim$70\%) of elliptical galaxies or early
type spirals  have an active galactic nucleus and $\sim$70\% of them
are LINERs. \textbf {We find a larger fraction of AGN in early
morphological types, as also found in the general population of
galaxies.}  Only 3\% of the AGN show the presence of broad lines (not a
single one can be classified as type 1 AGN).  This is a remarkable
result which is at odds with the unified model even if we
consider warped or clumpy tori. Finally, we interpret the large
fraction of AGN in isolated galaxies as the result of \textbf {secular
accretion.}

\end{abstract}

\keywords{ Galaxies: active - Galaxies: bulges - Galaxies: evolution -
Galaxies: formation}

\section{Introduction}

Along the last 25 years, many authors have studied the AGN environment
\citep{1982ApJS...50..517S, 1982ApJ...262...66S, 1984PhDT.........5D,
1985ApJS...57..643D, 1984ApJ...279L...5K, 1995IAUS..164..434D,
1996A&A...308..387D, La95, Dul99, Kron01, Kron02, 2001AJ....122.2243S,
2006A&A...451..809S,  2006ApJ...651...93K, 2006ApJ...639...37K,
2008RMxAC..32..150M}. The incidence of nuclear activity in galaxies
and their environment has become a topic of debate because there are
different mechanisms that can possibly trigger nuclear activity
depending on the galaxy's environment.  Interactions between galaxies
are well known to produce enhancement in star formation in galaxies
\citep{1984ApJ...287...95L, 1987AJ.....93.1011K,
1993AJ....106.1771K,Kron02, 2000ApJ...530..660B, 2007AJ....134..527W,
2007ApJ...660L..51L}. Others authors also have evidence for a
connection between circumnuclear starburst and AGN \citep[and
references therein]{2008RMxAC..32..139S}. There are also suggestions
for a connection between  interactions and nuclear no-thermal activity
specifically of type 2 \citep{Dul99,Kron01}. Other studies have dealt
with the AGN population of compact groups \citep{Mar10} as well as in
larger groups.

 The purpose of this project is to investigate the 
conditions to trigger AGN activity in different environments. In this
paper we study the incidence of activity in isolated galaxies.  In a
forthcoming paper \citep{HI13} we will
present the results of a survey of AGN in paired galaxies of similar
mass.

 It is important to study  galaxies in a restricted environment in 
order to elucidate what mechanisms could be determinant to trigger AGN
activity.  Isolated galaxies can  be defined as those systems that are
 formed in low galactic
density environments, but that evolved without  major interactions with other galaxies
of not only similar mass over the last 3 Gyr.  In this context, any
non-axisymmetric structures in these galaxies such as bars, tails,
plumes or stripping material must be the result of secular
evolution.\\

The study of truly isolated galaxies is thus fundamental to benchmark
the role of interactions in nuclear activity. Studies of field
galaxies  (e.g. Ho et al. 1997) cannot provide this information as
these samples may include galaxies that have undergone or are
undergoing an interaction.  

\textbf {This is the first paper of a series involving a self
consistent and homogeneous way to study nuclear activity in galaxies
in different environments}.  In the present work, we study the
incidence of nuclear activity in two samples of bona-fide isolated
galaxies using an efficient way to extract the stellar  contribution
(host spectrum). The purpose of this work is to have a well defined
sample of isolated galaxies with optical spectroscopic
characteristics that allow us to classify them according to their type
of activity. 
\textbf {Studying the incidence of the nuclear activity in
isolated galaxies alone is of great value to establish if AGN is a
common and/or persistent phenomenon even when strong tidal external
perturbations have not been present during the last few Gyr of galaxy
evolution. This would indicate that AGN activity can be triggered by secular
evolution processes in galaxies. Our results on this sample will be
further used as a benchmark to compare the  incidence of activity  in
a sample of isolated pair of galaxies \citep{HI13}. Our samples of
isolated and paired galaxies have been chosen to have consistent
properties with each other except for the presence of a companion.}

This paper is organized as follows. In \S 2 we describe our
samples. In \S 3 we present the data analysis and classification.
Results are given in \S 4 and \S 5 contains the discussion about the
possible mechanisms for developing an AGN in isolated galaxies.

\section{Characteristics of the Samples}

 \textbf {As stated above the so called ``field galaxies'' cannot be
used as a proper sample in the study of the properties of truly
isolated galaxies.  Therefore in this study we used two samples of
rigorously defined isolated galaxies: The photometric catalog of
isolated galaxies (CIG) by \citet{Kara73} and the northern isolated
disk galaxies compiled by \citet{Var04}.}

We take all available spectra from DR7 of SDSS
\citep{2009ApJS..182..543A} for the two samples. The spectra have a
wavelength coverage from 3800-9200\AA{} with  a resolution power of
1800-2200 and a signal-noise (S/N) \textgreater 4 per pixel at
g=20.2. The SDSS spectra were taken through a fiber aperture of 3
arcsec  in diameter (corresponding to 700 pc radios in average for our
samples). This means only that central regions of the galaxies were
observed.\\

\textbf {The (CIG) catalog contains 1051 galaxies. It is one of the
best sources of isolated objects.  The isolation criteria are still
used as the basis for new catalogs of  isolated galaxies
(e.g. \citet{2005A&A...436..443V, 2010ASPC..421...11K,
2010AJ....139.2525H, Co11, 2011AstBu..66....1K}). These isolation
criteria guarantee that the  galaxies have not experienced a major
merger/interaction over the last 3 Gyr.}

 The catalog was based on a visual search of northern-sky  galaxies
($\delta$ $\geq$ -3$^{o}$) with a magnitude limit of m$_{Zw}$ $\leq$
15.7 \textbf {and a range in z from $\sim$ 0.01 to 0.05}. Only objects
\textbf {which have} high galactic latitude are considered in order to
avoid galactic extinction\footnote[1]{\textbf {We know that our
results are insensitive to reddening because we use line emission
ratios which cancel this effect.}}  This sample, is reasonably
complete ($\sim$ 90\%) in the magnitude range 13.5$\leq$ m$_{Zw}$
$\leq$15.7 \citep{Her-Tol99}.  In this catalog a galaxy is considered
to be isolated when \textbf {it does not have a neighbor of similar
size (diameter $>$ $\frac{1}{4}$ of the target galaxy) within 20
diameters. This corresponds to a magnitude difference $\sim$ 3
(excluding any possible AGN luminosity contamination).  Considering a
``field" velocity of 150 km/s for a galaxy with diameter of 25 kpc, it
would require $\sim$ 3 Gigayears for a companion galaxy to abandon the
area enclosed in the 20 diameters of isolation criterion. A similar
time would be required to erase morphological perturbations due  to a
merger. This means that these galaxies have been unperturbed on
average by at least that time.} 
 
\textbf {The second sample we examined is that of northern isolated
disk galaxies compiled by \citet{Var04} wich originally contains 203
disk galaxies.  This sample considers different criteria for
isolation. In particular, it is based on the logarithmic ratio f,
between inner and tidal forces acting upon the candidate galaxy by a
possible perturber \citep[see][equation 3]{Var04}.  Only galaxies with
low f ratio (f $\leq$ -4.5) are considered as isolated because they do
not show signatures of any perturbation. They estimated that the
objects in their sample have not been affected by other galaxies
during the last 2 Gyr. It is important sample obtained with different
criteria to asses the reliability of our results.} 

 We analyzed 413 spectra for the CIG catalog and 100 for Varela's
sample from SDSS DR7.  After a meticulous inspection of the SLOAN
images, we excluded those spectra which: 1) Do not have the optic
fiber in the center of the galaxy (KIG 479 and KIG 237) or incomplete
spectra like in KIG 702 and KIG 479, 2) Are Blue Compact/H II Galaxies
(6 galaxies  in Varela's sample, 1 in CIG), 3) Galaxies showing traces
of interaction (tidal tails etc) or present a companion namely: KIG
349, KIG 439, KIG 468, KIG 634 and KIG 687
\citep{2006A&A...449..937S,2007A&A...472..121V} and 4) Galaxies that
did not achieve 3 $\sigma$ detection in all their line intensities
(37 galaxies of CIG sample).
 
\textbf {A visual inspection was performed for all galaxies to confirm
the morphological classification according to NED, SIMBAB and
HYPERLEDA.  In those few cases where an obvious misclassification was
present, the morphology was corrected by us.} In addition we found
that PGC 33255 could be part of a pair and excluded this galaxy of the
sample. Strangely enough, we found two elliptical galaxies with a very
blue compact core: PGC29177 and PGC43121 that show typical HII region
spectra, and were also excluded from our analysis. With this into
account, our spectroscopic sample from CIG consists of 367 galaxies
while the spectroscopic  sample from Varela consists of 93. Out of
these galaxies 18 and 10 respectively do not present emission
lines. In Table 1 we show the general statistics for both samples.

 We were careful to distinguish between intrinsic no emission and a
problem of detectability related to low S/N. For this purpose we set a
threshold of 10$^{38}$ erg s$^{-1}$ in H$\alpha$ luminosity.  The
galaxies below this threshold are the true no-emission objects with a
probability of being an AGN of less than 2\% and 4\% for the CIG and
Varela's samples respectively.  The distribution of morphology and
H$\alpha$ luminosity of our samples are presented in Figure 1. 

\section{Data Analysis and Nuclear Classification}

 We examined all of the spectra looking if emission lines were
present. Within the spectral range covered by the SDSS spectra we
searched  for H$\beta$, [OIII]$\lambda$5007\AA,[OI]$\lambda$6300\AA,
[NII]$\lambda\lambda$6548,6584\AA, H$\alpha$ and the two Sulfur
([SII]$\lambda\lambda$6717,6731\AA) lines.  In many cases the nebular
emission is very weak or it can be even diluted in the strong stellar
continuum of the galaxy. Since the integrated SDSS spectra  are
collected through 3 arcsecs fibers, they include not only the nuclear
emissions but also the surrounding stellar light coming from the host
galaxy.  This contamination turns out to be more significant at the
central parts of the galaxies and as the spheroidal/bulge component
becomes more relevant.  In fact it can even masks weak emission lines
such as those detected in galaxies with an AGN. Therefore to obtain a
reliable nuclear classification based on the emission lines it is
mandatory to subtract the stellar contribution. We applied the
principal component analysis (PCA)  method following
\citet{2005AJ....129.1783H} to subtract this contribution. We used
their first 8 eigenspectra from their low redshift range.  These
eigenspectra are the resulting eigenvectors of a PCA analysis applied
to a sample of high S/N spectra of non-emission galaxies. In addition,
as they pointed out, we included two more  components, an A star
spectrum accounting for the possible presence of post-starburst
features and a power-law to take into account of the possible
existence of a non-thermal component.  The analysis is performed for
all the spectra of our sample and it consists on a multiple regression
of each spectrum to a linear combination of the 8  eigenspectra plus
the two additional components. Previously to the fit each galaxy
spectrum was moved to zero redshift which is the one of the template
library.  We also masked all those regions where emission lines may
appear since the quality of the fit lies on the matching of the
continua. Once the regression is performed, the direct subtraction of
the resulting fit to the original (z=0) spectrum provides us with a
pure emission line spectrum where all the underlying absorption
components and eventually a non thermal component of the continuum are
removed.

\textbf {Line fluxes were calculated with Sherpa software
(http://cxc.cfa.harvard.edu/sherpa/) (which comes in the CIAO
distribution, http://cxc.harvard.edu/ciao/).  Sherpa reads the data
and evaluates a given model on this data set. Then, it varies the free
parameters to minimize a statistical goodness function to  obtain the
best set of parameters that fit the data. In our case our model is
composed only by Gaussians. The width and velocity of all the
detected lines was constrained to the same value in our fits.
Therefore we have as free parameters these two quantities and measured
the intensity of each emission line. For those objects where a broad
component was required in addition to the narrow one, an individual
broad Gaussian was fitted with fully independent free parameters (see
Figure 2 for fit examples). }  
 
We have used the Baldwin, Phillips and Terlevich optical diagnostic
diagrams \citep{1981PASP...93....5B, 1987ApJS...63..295V} to
separate star-forming galaxies from active galactic nuclei
(AGN). Line ratios adopted were ([O III]/H$\beta$), ([N
II]/H$\alpha$), ([S II]/H$\alpha$) and ([O I]/H$\alpha$). With these
ratios we produced the ([O III]/H$\beta$) vs ([N II]/H$\alpha$)
diagram (hereafter [N II] diagram), the ([O III]/H$\beta$) vs ([S
II]/H$\alpha$) [S II] diagram and the ([O III]/H$\beta$) vs ([O
I]/H$\alpha$) [O I] diagram. Line ratios with their errors and AGN
type are presented in Tables 6 and 7 for CIG and Varela's samples
respectively.  \textbf {The last column in these tables gives the
nuclear type. We have used the quantitative definition by \citet
{1992MNRAS.257..677W} to measure the contribution of  a broad
component when present, and thus establish the Seyfert type. }

 We used two demarcations to identify the galaxy type.  The first one
is the theoretical division line proposed by \citet{Ke01} (K01 line
hereafter) where objects above this line on the diagrams have nuclear
activity.  The second one is the empirical line derived by
\citet{Ka03} (K03 line hereafter), this empirical line is based on the
location of star-forming galaxies  for the SDSS. Galaxies that lie
between this two lines are called composite (or transition)
objects. This objects need non-thermal processes to produce the line
ratios (in addition to star forming processes).  Therefore, we will
take these objects as AGN in this work. Further support comes from the
work done by \citet{Trouille11} where they found that  composite
galaxies on BPT [N II] diagram are X-ray hard sources and have a high
X-ray luminosity to total infrared luminosity.

We have performed the analysis of the CIG catalog studying the effects of galactic morphology.
Figure 3 shows a histogram  of morphological  types for both
samples. It is clear that around 70 \% of the galaxies in the CIG have
morphologies between Sb and Sc. This result is consistent with a work
done by \citet{2010AJ....139.2525H} for other isolated galaxy sample.
The fraction of AGN for each morphological type in both samples is
shown in Figure 4. We can see that within the large statistical errors
the comparison  is valid. 

\section{Results}

 Our main results are summarized in Tables 2 to 5, and also in Figures
4 to 8.

In Table 2 we show numbers and percentages of  activity and
morphological classes for the CIG sample. Column 1 presents the
morphological type, the total number of galaxies is listed in column
2.  Columns 3 and 4 list the number of galaxies with H II and
Composite (hereafter Comp) activity and their percentages,
respectively. Column 5 contains the number of galaxies in the AGN+Comp
region and their percentages. In column 6 we present the total number
of galaxies including those with no-emission lines. Columns 7, 8  and
9 show the number and percentages for H II, Comp and AGN+Comp activity
respectively. In Table 3 we present the same data as in Table 2 for
Varela's sample.

In Figure 5 we show the [N II] BPT diagram. All galaxies in the
composite region are taken as AGN as argued in \S 3. 

 We can clearly see that the incidence of nuclear activity is
statistically higher in early type galaxies and decreases gradually as
we go to late types in this subsample of emission lines galaxies only.
The dependence on morphology is quite significant: going from 70 \% in
E galaxies to 10 \% in Sm.  The incidence of nuclear  starburst
activity decreases in the opposite sense: going from 90 \% for Sm to
30 \% for E. Meanwhile, if we consider the total sample (including
galaxies without emission lines) the  incidence gives a flatter
distribution from E to Sb and decreases only for late types (compare
columns 5 and 9 in Tables 2 and 3). This is clearly observed in Figure 4.

In Figure 6, we show the position of our objects in the BPT [N II]
diagram for Varela's sample. In this sample there are only 4
ellipticals, whereas the majority (27) are Sc. Considering that early
types have a larger incidence of AGN, this can explain the fact that
we find a total incidence of AGN  of only 31 \% as compared to 43 \%
in the CIG sample.
 
In Figure 7 we present the [S II] diagnostic diagram for the CIG
sample. This diagram is useful to separate between Seyferts and
LINERs.

In Figure 8. We show yet another diagram [O I]. This diagram differs
only slightly from the one in  Figure 7. The main difference is in the
relative proportion of Seyferts and LINERS except for ellipticals. We
point out that the [O I] line is weaker than the  [S II] line and thus
the mean error is large. 

 In Table 4. We quantify the incidence of AGN activity and morphology
distribution derived from the [S II] and [O I] diagrams for CIG
sample.  Columns 1 and 2 are the morphological type and the number of
galaxies in each type.  In column 3 we list the number of AGN and the
percentage in parentheses. The number of Seyfert galaxies and the
percentage are  listed in column 4. The number of LINERs is in column
5. Columns from 6 to 9 contains the same data of columns of 2 to 5 but
this time for the [O I] diagram. Table 5  is the analogous of Table 4
for Varela's sample.
 
\textbf {From these Tables we can confirm the result that the incidence of AGN
activity is higher for early type galaxies in the sample with emission
lines only. When all galaxies are considered the distribution flattens
as found before for the CIG sample (see Fig.4). Again, the incidence of Seyfert
galaxies is more frequent in morphologies from Sa to Sc, but specially
in Sb.}   

\textbf {In Tables 4 and 5 we show the data for the [S II] and [O I]
diagrams. The AGN fractions are different from those in Tables 2 and
3.  The reason is that we do not consider composite nuclei in the [S
II] and [O I] diagrams. Unfortunately models by \citet{Ka03} for these
particular diagnostic diagrams ([S II] and [O I]) are not
available. In consequence we only take into account the AGN activity
with the Kewley limit for self-consistency.}

 We want to point out to recent models developed by
\citet{2006MNRAS.371..972S}. These models predict lower values for the
ratios of [N II]/H$\alpha$ and [O III]/H$\beta$ for AGN. In a Figure
analogous to Fig. 5 the models by \citet{2006MNRAS.371..972S} would
produce an even larger zone of composite objects.

\textbf {The most remarkable result is the absence of type 1 AGN for both
samples. In the CIG sample there are 12 galaxies which show lines with
a broad component. Five of these galaxies (all of them classified as
Seyfert nuclei) have a clear broad component, three are Sy 1.5 (KIG
214, KIG 747, KIG 1008), one is Sy 1.8 (KIG 749) and one is Sy 1.9
(KIG 349). Other three objects (KIG 204, KIG 603 and KIG 605) are classified as LINERs. 
For two additional AGN (KIG 553, KIG 591) a Seyfert/LINER classification was not possible. 
The remaining two objects are HII galaxies with broad components. 
Broad components in HII regions are rare but have indeed been found \citep [e.g.][]{2009A&A...500..817B}.}

\textbf {In Varela's sample there is only one 1.8 type Seyfert galaxy (PGC48521). This object was classified as Sy 1 in SIMBAB and Sy 1.9 in NED.
 It has been shown that a few galaxies can vary their type with time \citep [e.g.][]{2008A&A...486...99S, 2012ApJS..202...10S}. 
Whether this is the case, or this object was misclassified in SIMBAD has little relevance for our general
conclusions.} 

\textbf {To be conservative, we consider these 13 galaxies as the fraction
of possible AGN with broad components (including the two HII galaxies to allow for any possible misclassification). 
These numbers indicate that the fraction of type 1 objects is $\lesssim$ 3\%.}

\section{Discussion}

The large number of galaxies of all morphological types permits us to
quantify the link between morphology and nuclear activity.  Our
results indicate a close link between these two properties. This implies that any result of the
incidence of activity without this consideration reflects the
particular morphological distribution of the sample and therefore is
not reliable. Our sample includes for the first time a statistically
significant number of isolated early type galaxies (E+S0).

 \textbf {We found that Elliptical and SO galaxies have the highest incidence of
nuclear activity in isolated environments when only galaxies with emission lines are considered (a similar result was found by \cite{Var04, Co11, Sa12}).
 However, when the total sample is taken into account (including galaxies without emission lines) 
these apparent excess disappears and all early types (including Sa and Sb types) have similar fractions (see Fig. 4). 
This important difference could be found thanks to the large number of elliptical and spheroidal galaxies in our sample.
 This results are consistent with those found for ``field" galaxies \citep {1980A&A....87..142H, 1983ApJ...269..466K, Ka03, 2003ApJ...597..142M}. 
Finding the same trend between isolated and field galaxies could be expected if we consider that AGN require SMBHs and black holes are correlated with bulges. 
However, the large number of AGN among isolated galaxies is of great importance and indicates that secular evolution processes can trigger/maintain low luminosity AGN activity (see below).}

We consider as secular evolution the following mechanisms capable of driving gas into the nuclear region: 1) Minor mergers (luminosity
ratio larger than 10 in our sample); 2) Dark matter accretion; 3)Non-axisymmetric gravitational perturbations. 

With respect to dark matter, \citet{2010MNRAS.404L...6H} showed from
an analytical treatment of the accretion rate that, for the largest
black hole masses of quasi-stellar objects ($>$10$^{9}$ M$_{\odot}$),
the runaway regime would be reached on time-scales wich are shorter
than the lifetimes of the halos in question for central dark matter
densities in excess of 250 M$_{\odot}$/pc$^{3}$. This limit scales
inversely with the black hole (BH) mass.    
 
\textbf {The most common non-axisymmetric internal potential is due to
the presence of a bar. However in  the particular case of barred
galaxies it has been shown that most probably bars do not enhance
nuclear activity \citep{1995ApJ...438..604M, 1997ApJ...487..591H,
2012arXiv1203.1693L}. However the samples  used in those studies were
not rigorously isolated and thus the effects of the environment cannot
be disentangled from those of the bar. The samples used in this work
provide the opportunity to perform a rigorous test of the effect of a
bar. This can be achieved due to both the selection criteria and  the
quality of the data. A detailed analysis of the bar fraction  requires
a deep photometric study. This analysis will be presented in a
forthcoming paper \citep{Her-Tol13}.}

 \textbf {We note that although a large fraction of isolated galaxies are active, their SMBH has not grown
significantly over the last 3 Gyr. Given that most of our  galaxies
are representative of the low luminosity end of AGN, the mass
accretion rate should be in the range of $10^{-5}-10^{-3}$
M$_{\odot}$/yr, and the radiative efficiency $\eta$ should be
significantly smaller than 10 \% \citep{2003ASPC..290..379H,
2009ApJ...699..626H}. Such low efficiencies are predicted  for low
luminosity AGN \citep{2008NewAR..51..733N}.  Assuming the AGN in our
sample have accreted at a constant rate over the last 3 Gyr the growth
of their SMBH ranges between $~10^{4} - 10^{6}$ M$_{\odot}$.  Then, it
is clear that isolated galaxies in poor environments have
failed to accrete enough material  (at least during the last Gyr) to present higher luminosities and significant
black hole growth.  Thus, our results support a hierarchical scenario
in which the environment is crucial to determine properties such as
luminosity, mass, and central SMBH mass, fulfilling the expectations
of the downsizing for SMBH growth \citep {2010hsa5.conf..337P}.} 

The spiral isolated galaxies will not migrate from the blue to the red sequence
since feedback is not efficient in these faint AGN
\citep{2007ApJ...659.1022K}.  The later result supports again that
secular evolution in these galaxies is the important mechanism to
establish the bulge-black hole relation.  This is in contrast to the
case of massive galaxies transitioning from the blue to the red branch
of the color-color diagram which require a major merger followed by a
substantial feedback in the QSO phase. 

 On the other hand our isolated ellipticals are already in the
red-branch of galaxies that probably have experienced a major  merger
in the distant past.  The fact that essentially all of them are AGN
may simply reflect the fact that it is easier to drive gas to the
center of spheroidal systems.  The remanent inter stellar medium (ISM)
in these galaxies is  typically in the range (10$^{6}$-10$^{7}$
M$_{\odot}$). Therefore, they contain enough gas to power their SMBH
over the last 3 Gyr. This does not exclude, however, the possibility
of an external supply of material as suggested by several authors
\citep{1992ApJ...401L..79B, 2000ApJS..127...39C,
2006MNRAS.366.1151S}.  The large fraction of AGN in these galaxies
suggests that the presence of a large bulge facilitates the mass
in fall to the center.  We note, however, that a small fraction of this
LINERs could be fake AGN (``retired galaxies'',
\citet{2011MNRAS.413.1687C} ).

Finally the absence of type 1 AGN in these samples of isolated
galaxies is remarkable and at odds with a simple interpretation of the
unified model (UM). There is not a single type 1.0 AGN among the 175
active galaxies in our samples. The fraction of types 1.5-1.9 is less
than 3 \% in both isolated galaxy samples. 

\textbf {All these results indicate that the presence of AGN activity
is a common phenomenon in galaxies independently of the
environment. This is an important  result as traditionally it has been
assumed that an external perturbation is required to induce nuclear
activity. Our results indicate that a low  luminosity AGN phase is a
part of the secular evolution of a large number of galaxies. These
findings are consistent with those by \citet[and references therein]
{1997ApJS..112..315H, 1997ApJ...487..568H, 2002ASPC..284...13H}.
However, in those studies it was impossible to disentangle the
environmental effects from those of internal galactic evolution, given
that the isolation  history was not known a priory in their
samples. Our results do not deny  the possibility that external
perturbations may enhance the frequency of nuclear activity among
galaxies, as has been suggested by previous studies \citep{Dul99, Kron02,
Kron03, Ro09, 2010MNRAS.407.1514E, 2011MNRAS.418.2043E}. The
effect of a strong gravitational interaction will be studied in a
forthcoming paper  where the incidence of AGN in a sample of isolated
close pairs of similar mass galaxies will be analyzed. We also note that the lack
of high luminosity AGN in  our samples points towards a dependence
between environment/interactions and AGN luminosity. In this scenario,
the extremely low fraction of type 1 AGN  can be understood if a BLR
(broad line region) can be formed only at higher accretion
rates/luminosities
\citep{2000ApJ...530L..65N,2009ApJ...701L..91E}. The appearance of a type 1
nucleus may be delayed by as much as 1 Gyr, as required by the
evolutionary model proposed by \citet{2007ApJ...659.1022K}, where an
interaction triggers first a circumnuclear starburst, and
subsequently non-thermal nuclear activity. For the brightest end of  nuclear
activity a similar evolutionary trend is possible (from ULIRGs to luminous quasars). In this case a major merger would
be required, affecting the overall properties of the host galaxy and
moving it to the  blue  branch of the color-color diagram.}

\textbf {If AGN in isolated galaxies have low accretion rates, low
efficiencies, low luminosities and almost a complete absence of broad
lines in their spectra, it is probable that the BLR under these
circumstances is not even able to form. This is in
accordance with the result by  \citet {2003ApJ...583..632T,
2003ASPC..290...31T} for the absence of broad components in polarized
light for $\sim$50$\%$ of  the galaxies in his sample. Several
studies show evidence that the Sy2s with and without broad lines in
polarized lines  (in other words with and without a hidden BLR) are
truly different in other respects as well  \citep [e.g.][]{2001RMxAA..37....3G, 2001A&A...366..765G, 2012arXiv1209.0274B}.}

\acknowledgments{FJHI acknowledges a graduate student scholarship from
CONACYT. DD acknowledge support from grant IN111610 PAPIIT, UNAM.}

\clearpage
\begin{figure}
\includegraphics[scale=1.2]{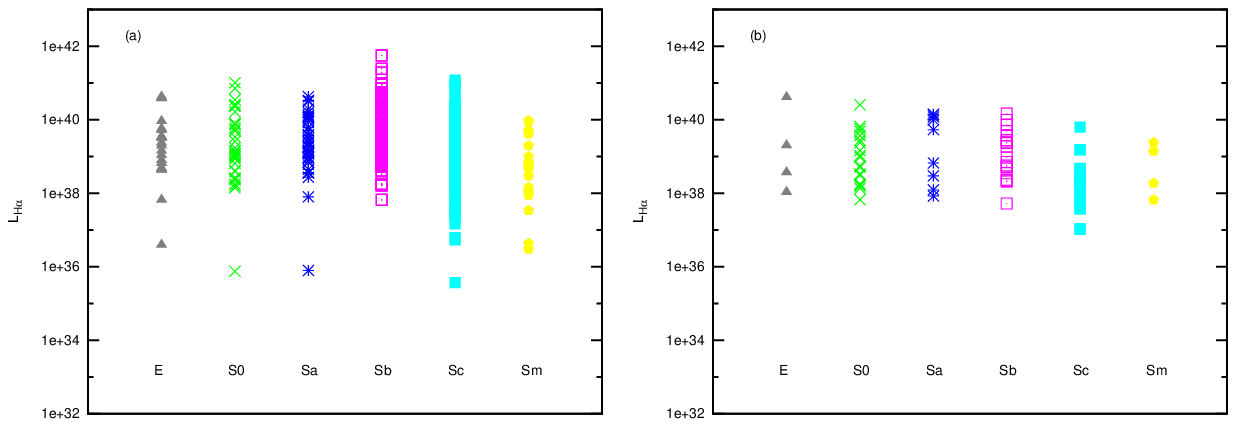}
\caption{Morphological distribution of H$\alpha$ luminosity for (a)
CIG sample and (b) Varela's sample. Mean values of AGN H$\alpha$
luminosity for CIG and Varela's samples are \emph{L$_H$$_\alpha$ = 9.2
$\times$ 10$^{39}$ erg s$^{-1}$} and \emph{L$_H$$_\alpha$ = 1.8
$\times$ 10$^{39}$ erg s$^{-1}$} respectively  which correspond to Low
Luminosity AGN.}
\end{figure}

\clearpage
\begin{figure}
\includegraphics[scale=0.4]{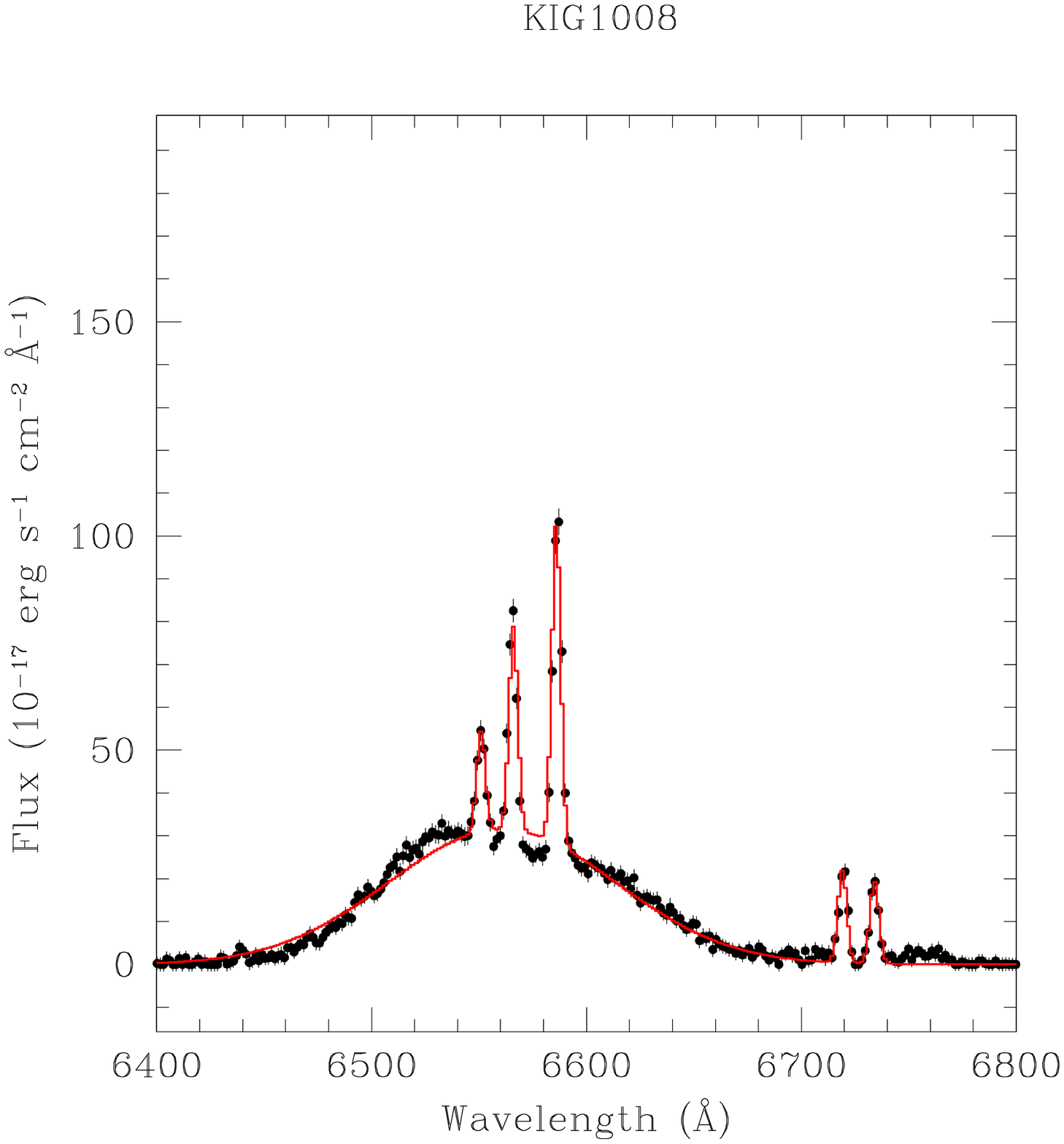}
\includegraphics[scale=0.4]{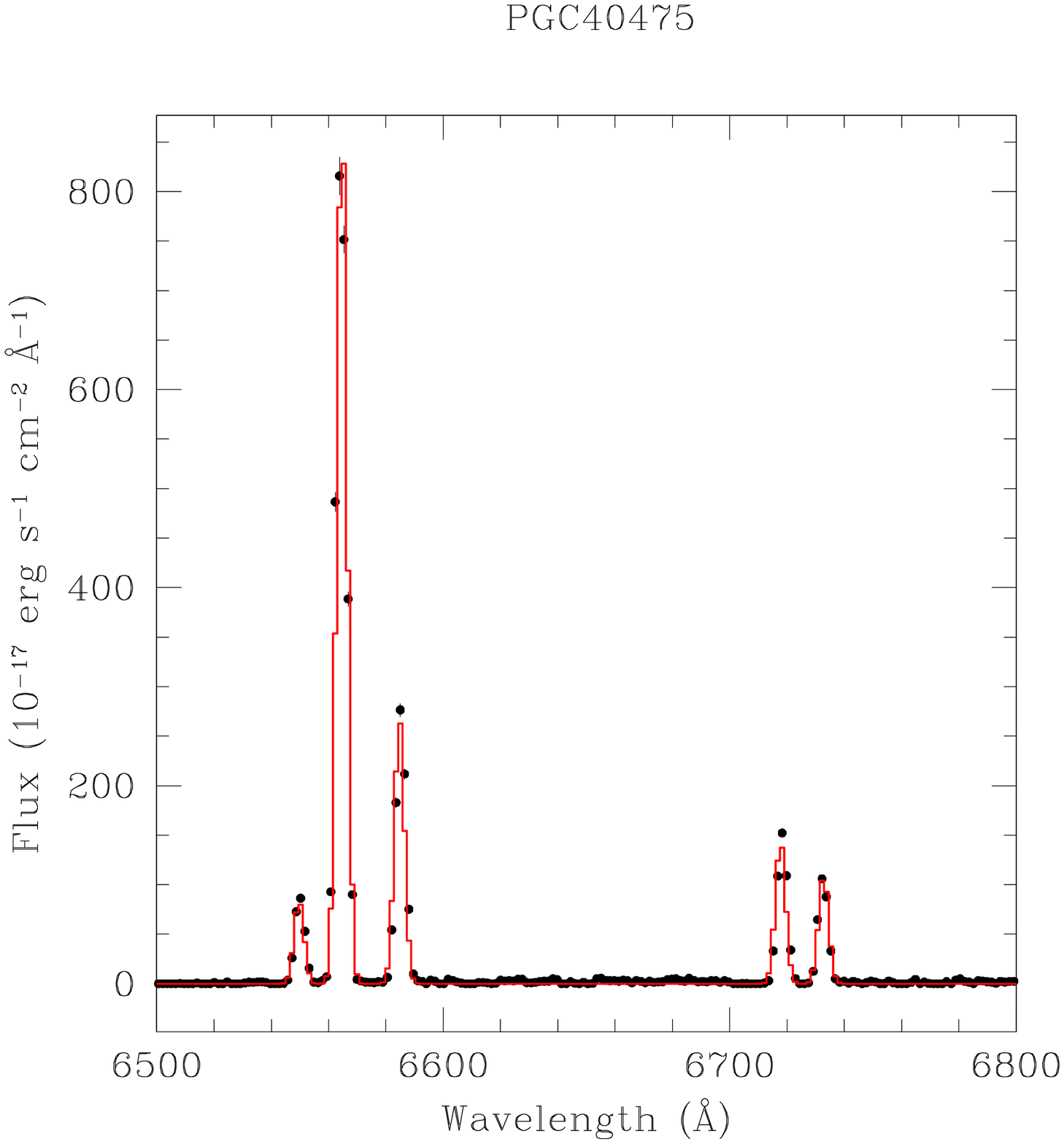}
\caption{Examples of fits in H$\alpha$ region for galaxies with and
without broad line respectively. Black data points denote the spectrum
with the stellar contribution subtracted and the red line shows the
fit.}
\end{figure}

\begin{figure}
\includegraphics[angle=0,scale=1]{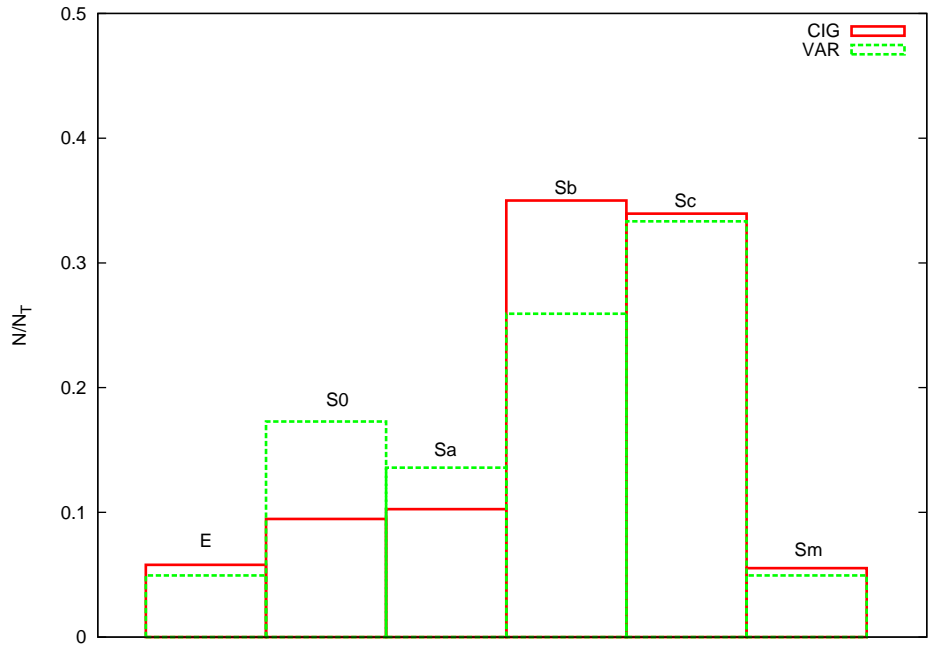}
\caption{Histogram of our samples with morphological type
separation. Continuous red line correspond to CIG sample and dashed
green line to Varela's sample.}
\end{figure}

\begin{figure}
\includegraphics[angle=0,scale=1]{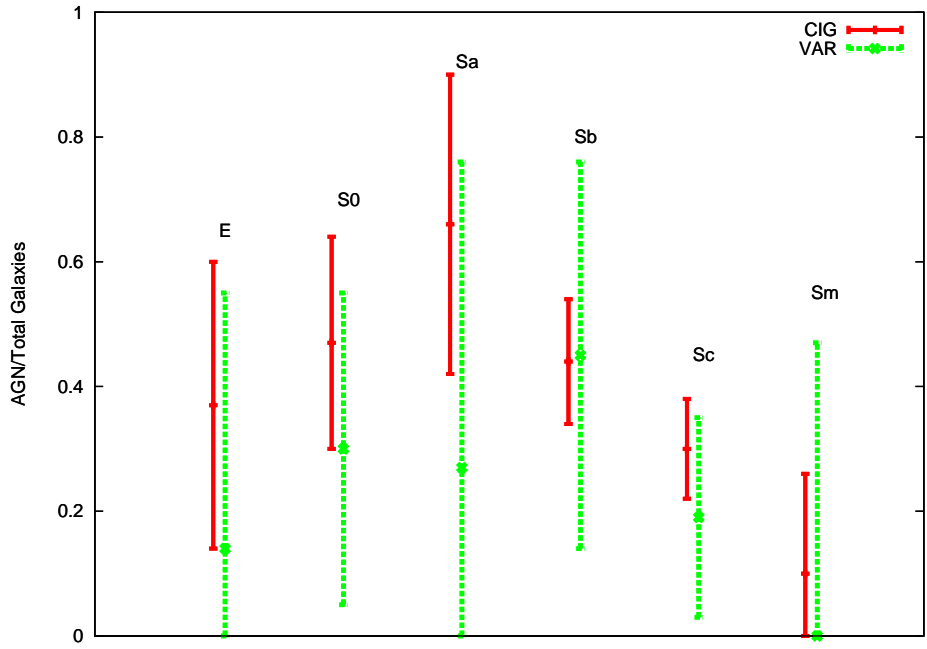}
\caption{Statistical comparison between isolated samples. Lines labels
like in Figure 3.}
\end{figure}

\begin{figure}
\includegraphics[scale=1.4]{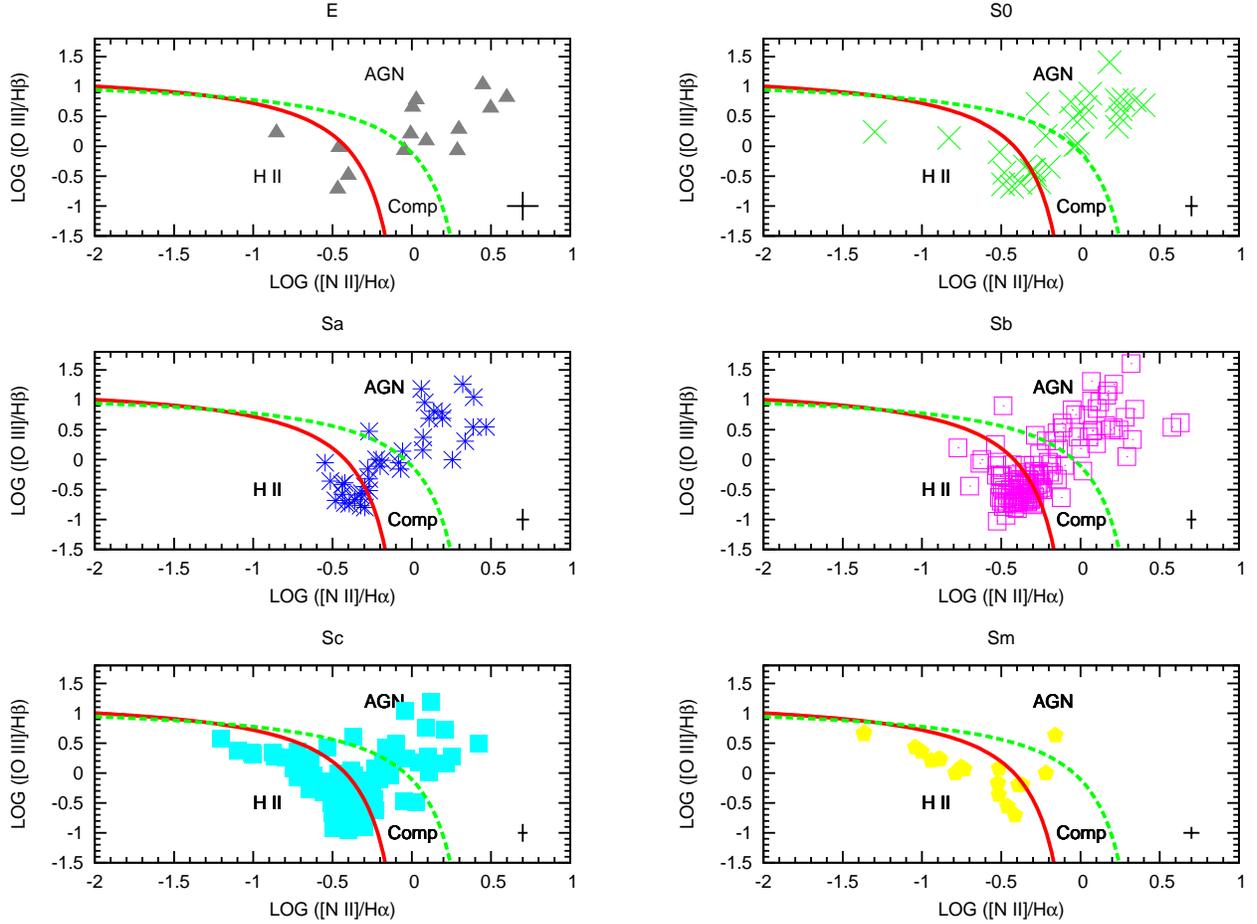}
\caption{The [N II] diagnostic diagrams for CIG sample with different
morphologies. This diagram separates between tree different kind of
activity in galaxies like AGN, Composite and H II like region
galaxies. The green dashed line (Ke01) separates galaxies with a AGN
from Composite (AGN+Starburts activity). Continuous red line (Ka03)
divides pure star forming galaxies from AGN-starburts composite
objects. Elliptical galaxies are shown as filled gray triangles,
lenticular as  green crosses, Sa as blue asterisks, Sb as pink empty
squares, Sc as filled cyan squares and Sm as filled yellow
pentagons. The cross at the lower right part of the diagram is the
mean error in the line ratios.}
\end{figure} \clearpage

\begin{figure}
\includegraphics[scale=1.3]{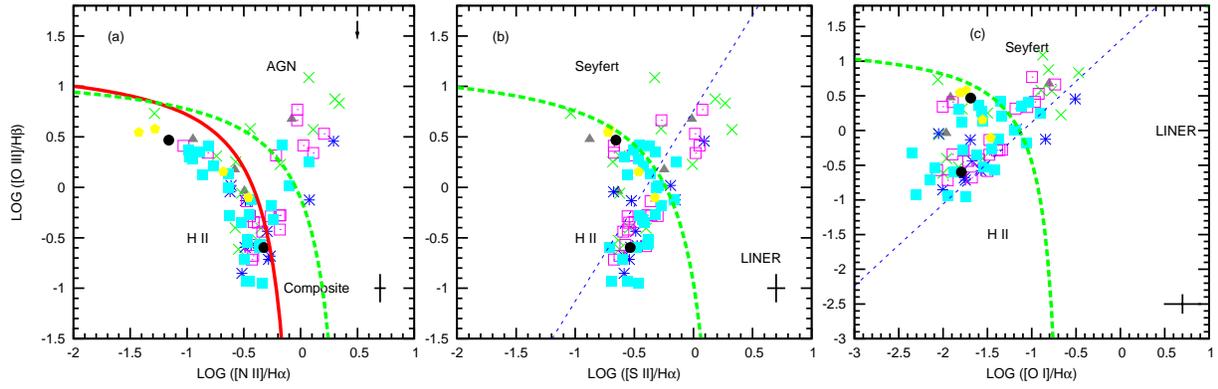}
\caption{(a) The [N II], (b) [S II] and (c) [O I] diagnostic diagrams
for Varela's sample. Labels like on Figure 5.  The low incidence of
nuclear activity are present on Sc and Sm types. Filled circles
represent objects which cannot be classified.}
\end{figure} \clearpage

\begin{figure}
\includegraphics[scale=1.4]{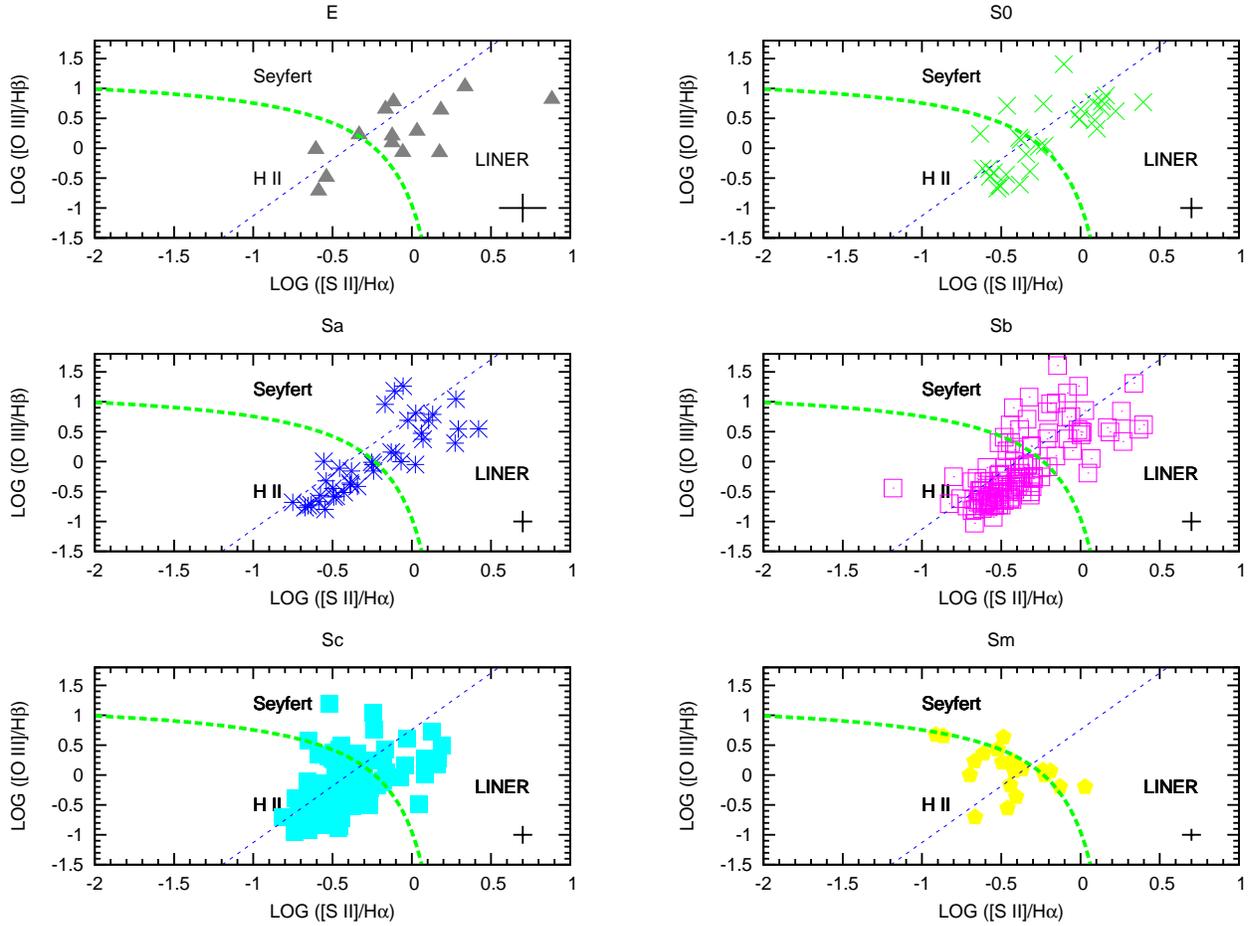}
\caption{The [S II] diagnostic diagram for CIG sample. Blue dashed
line represents Seyfert/LINER line and others labels like in Figure
5.}
\end{figure} \clearpage

\begin{figure}
\includegraphics[scale=1.4]{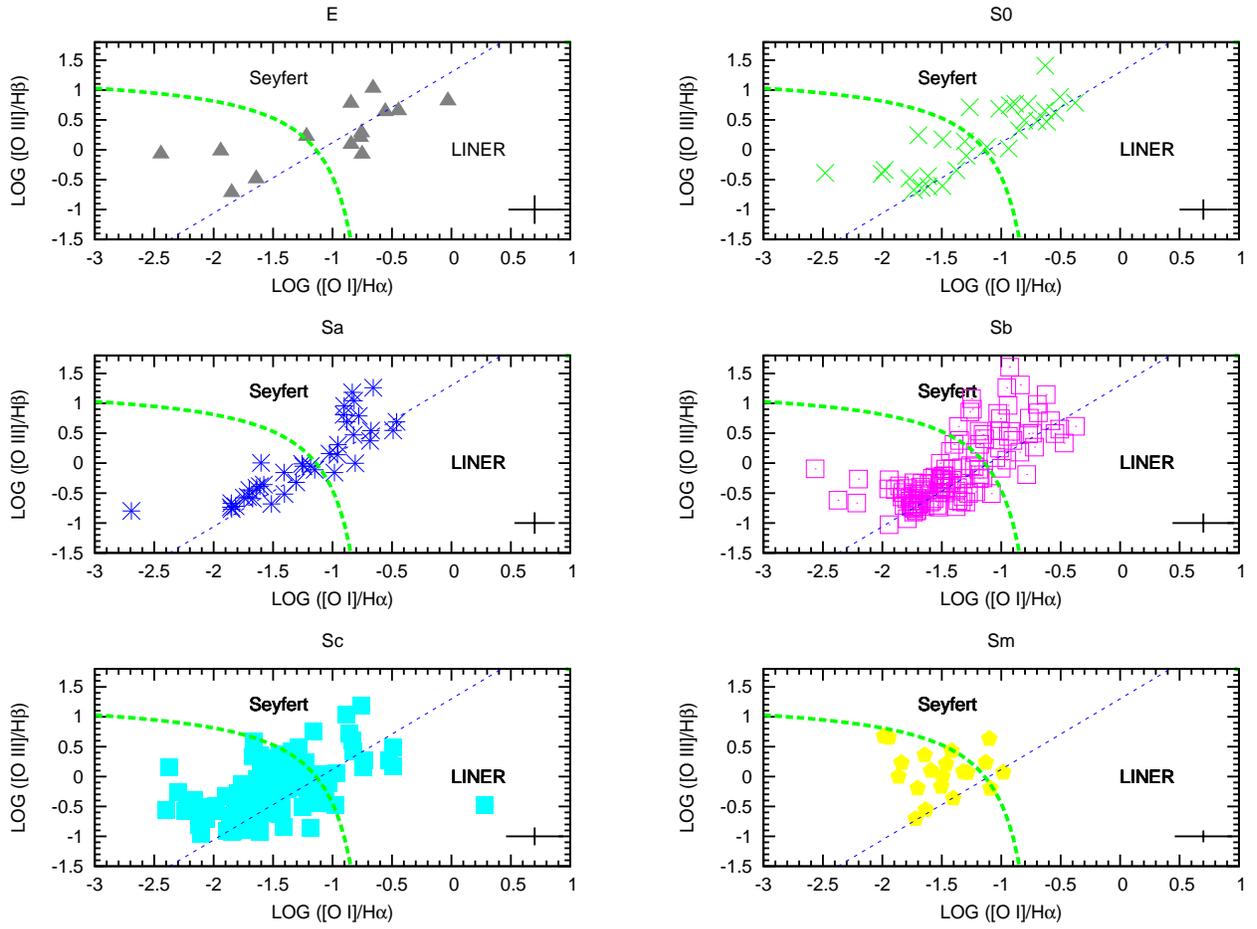}
\caption{The [O I] diagnostic diagram for CIG sample. Labels like on
Figure 7.}
\end{figure} \clearpage

\clearpage
\begin{deluxetable}{cccccccc} 
\tabletypesize{\footnotesize}
\tablecolumns{8} 
\tablewidth{0pt} 
\tablecaption{General statistics of both Isolated galaxy samples.}  
\tablehead{ \colhead{Sample}
&  \colhead{Total}&  \colhead{No Emission}      &  \colhead{Excluded}
& \colhead{Fitted} 	 & \colhead{AGN+Comp} 	 & \colhead{H II}
& \colhead{AGN Type 1}	} 
\startdata
\hline CIG & 413 & 18(4\%) & 47(11\%) & 348(84\%) & 150(36\%) &
198(48\%) & 12(3\%)\\ VAR & 100 & 10(10\%) & 7(7\%) & 83(83\%) &
26(26\%) & 57(57\%) & 1(1\%)\\ TOTAL & 513 & 28(5\%) & 54(11\%) &
431(84\%) & 176(34\%) & 255(50\%) & 13(3\%)\\ 
\hline
\enddata
\end{deluxetable}

\clearpage

    \begin{table}\centering \small
    \caption{Morphology distribution and incidence of nuclear activity
for the CIG sample derived from [N II] BPT diagnostic diagram.}
    \begin{tabular}{l l cccc l cccc} \hline\hline &
\multicolumn{4}{c}{Galaxies with Emission lines}
&&\multicolumn{4}{c}{Total sample}\\ \cline{2-5} \cline{7-10} M.T.&
Total & H II & Comp & AGN+Comp && Total & H II & Comp & AGN+Comp \\ E
& 14 & 4(29\%) &  1(7\%) &  10(71\%)  && 23 & 4(17\%) &  1(4\%)  &
10(43\%)  \\  S0   & 30 & 9(30\%) &  5(17\%) & 21(70\%)  && 38 &
9(24\%) & 5(13\%)   & 21(55\%)  \\  Sa & 38 & 13(34\%) & 8(21\%) &
25(66\%)  && 38 & 13(34\%) & 8(21\%)  & 25(66\%)  \\  Sb   & 122 &
67(55\%) & 20(16\%) &  55(45\%)  && 124 & 67(54\%) & 20(16\%)  &
55(44\%)  \\  Sc     & 124 & 87(70\%) & 22(18\%) &  37(30\%)  && 124 &
87(70\%) & 22(18\%)  &  37(30\%)  \\  Sm   & 20 & 18(90\%) & 1(5\%) &
2(10\%)  && 20 & 18(90\%)&  1(5\%) & 2(10\%)  \\     \hline Total   &
348 & 198(57\%) & 57(16\%) & 150(43\%)  && 367 & 198(54\%)&  57(16\%)
& 150(41\%)  \\ \hline  \hline 
      \end{tabular}
      
      \end{table}

\begin{table}\centering \small
      \caption{Morphology distribution and incidence of nuclear
activity in Varela's sample derived from [N II] BPT diagnostic
diagram.}    
      \begin{tabular}{l l cccc l cccc} \hline\hline &
\multicolumn{4}{c}{Galaxies with Emission lines}
&&\multicolumn{4}{c}{Total sample}\\ \cline{2-5} \cline{7-10} M.T.&
Total & H II & Comp & AGN+Comp && Total & H II & Comp & AGN+Comp \\ E
& 4 & 3(75\%) &  0(0\%) &  1(25\%)  && 7 & 3(43\%) &  0(0\%)  &
1(14\%)  \\  S0   & 14 & 8(57\%) &  1(7\%) & 6(43\%)  && 20 & 8(40\%)
& 1(5\%)   & 6(30\%)  \\  Sa & 11 & 8(73\%) & 1(9\%) & 3(27\%)  && 11
& 8(73\%) & 1(9\%) & 3(27\%) \\  Sb   & 21 & 11(52\%) & 4(19\%) &
10(48\%)  && 22 & 11(50\%) & 4(18\%)  &  10(45\%)  \\  Sc     & 27 &
22(81\%) & 3(11\%) &  5(19\%)  && 27 & 22(81\%) & 3(11\%) &  5(19\%)\\
Sm   & 4 & 4(100\%) & 0(0\%) & 0(0\%)  && 4 & 4(100\%) & 0(0\%) &
0(0\%)  \\     \hline Total   & 81 & 56(69\%) & 9(11\%) & 25(31\%)  &&
91 & 56(62\%)&  9(10\%) & 25(27\%)  \\ \hline  \hline 
      \end{tabular}
      \end{table}

\begin{table}\centering \small
      \caption{Incidence type of nuclear activity for different
morphologies from [S II] and [O I] diagnostic diagrams for CIG
sample.}    
      \begin{tabular}{l l cccc l cccc} \hline\hline &
\multicolumn{4}{c}{ [S II] Diagram} &&\multicolumn{4}{c}{[O I]
Diagram}\\ \cline{2-5} \cline{7-10} M.T.& Total & AGN & Seyfert &
LINER && Total & AGN & Seyfert & LINER \\ E & 14 & 11(79\%) & 3(27\%)
& 8(73\%) && 13 & 10(77\%) & 3(30\%) & 7(70\%)\\ S0 & 30 & 15(50\%) &
3(20\%) & 12(80\%) && 28 & 16(57\%) & 12(75\%) & 4(25\%)\\ Sa & 38 &
15(39\%) & 3(20\%) & 12(80\%) && 36 & 17(47\%) & 11(65\%) & 6(35\%)\\
Sb & 122 & 31(27\%) & 15(48\%) & 16(52\%) && 120 & 39(33\%) & 30(77\%)
& 9(23\%)\\ Sc & 124 & 21(17\%) & 8(38\%) & 13(62\%) && 123 & 22(18\%)
& 11(50\%) & 11(50\%)\\ Sm & 20 & 6(30\%) & 1(17\%) & 5(83\%) && 20 &
3(15\%) & 2(67\%) & 1(33\%)\\ \hline Total & 348& 99(28\%)& 33(33\%)&
66(67\%)  && 340 & 107(31\%) & 69(64\%) & 38(36\%)\\ \hline  \hline 
      \end{tabular}
      \end{table}
\begin{table}\centering \small
      \caption{Incidence type of nuclear activity for different
morphologies from [S II] and [O I] diagnostic diagrams for Varela's
sample.}    
      \begin{tabular}{l l cccc l cccc} \hline\hline &
\multicolumn{4}{c}{ [S II] Diagram} &&\multicolumn{4}{c}{[O I]
Diagram}\\ \cline{2-5} \cline{7-10} M.T.& Total & AGN & Seyfert &
LINER && Total & AGN & Seyfert & LINER \\ E & 4 & 2(50\%) &  0(0\%) &
2(100\%)  && 4 & 1(25\%) &  1(100\%)  &  0(0\%)  \\  S0   & 14 &
6(43\%) &  2(33\%) & 4(67\%)  && 14 & 6(43\%) & 5(83\%)   & 1(17\%)
\\  Sa & 11 & 3(27\%) & 0(0\%) & 3(100\%)  && 11 & 2(18\%) & 0(0\%) &
2(100\%) \\  Sb   & 21 & 5(24\%) & 1(20\%) &  4(80\%)  && 21 & 6(29\%)
& 6(100\%)  &  0(0\%)  \\  Sc     & 27 & 5(19\%) & 3(60\%) &  2(40\%)
&& 27 & 5(19\%) & 4(80\%) &  1(20\%)\\  Sm   & 4 & 0(0\%) & 0(0\%) &
0(0\%)  && 4 & 0(0\%) & 0(0\%) & 0(0\%)  \\     \hline Total   & 81 &
21(26\%) & 6(29\%) & 15(71\%)  && 81 & 20(25\%)&  16(80\%) & 4(20\%)
\\ \hline  \hline 
      \end{tabular}
      \end{table}

\clearpage

\begin{deluxetable}{ccccccc} \tabletypesize{\footnotesize}
\tablecolumns{7} \tablewidth{0pt} \tablecaption{Logarithm intensity
ratios with their errors and AGN type for the CIG sample.}
\tablehead{ \colhead{Object}			 &
\colhead{M.T.\,\tablenotemark{a}}  	 &  \colhead{LOG([O
III]/H$\beta$)}      &  \colhead{LOG([N II]/H$\alpha$)} 	 &
\colhead{LOG([S II]/H$\alpha$)}      & \colhead{LOG([O I]/H$\alpha$) }
& \colhead{Type} } \startdata \sidehead{Elliptical} \hline KIG378
&	E	&	0.2831	$\pm$	0.1763	&	0.2970	$\pm$
0.0806	&	0.0332	$\pm$	0.1534	&-0.7559$\pm$	0.3383&LINER\\
KIG393	&	E-SO	&	-0.0234	$\pm$	0.0074	&
-0.4586	$\pm$	0.0066	&	-0.6059	$\pm$	0.0091	&-1.9412$\pm$
0.0287&HII	\\ KIG437	&	E-SO	&	0.0909	$\pm$
0.1939	&	0.0919	$\pm$	0.0589	&	-0.1233	$\pm$	0.1182
&-0.8432$\pm$	0.2380&LINER	\\ KIG462	&	E	&
-0.7172	$\pm$	0.0384	&	-0.4667	$\pm$	0.0098	&
-0.5874	$\pm$	0.0155	&-1.8481$\pm$	0.0740&HII	\\ KIG555
&	E	&	0.2234	$\pm$	0.0401	&	-0.8536	$\pm$
0.0429	&	-0.3324	$\pm$	0.0298	&-1.2193$\pm$	0.0865&HII
\\ KIG556	&	E	&	0.8151	$\pm$	0.5101	&
0.5997	$\pm$	0.2621	&	0.8841	$\pm$	0.2651	&-0.0295$\pm$
0.4288&LINER	\\ KIG582	&	E	&	1.0277	$\pm$
0.6437	&	0.4476	$\pm$	0.1483	&	0.3357	$\pm$	0.2130
&-0.6587$\pm$	0.5464&LINER	\\ KIG595	&	E	&
-0.0732	$\pm$	0.2877	&	-0.0465	$\pm$	0.2346	&
-0.0569	$\pm$	0.3745	&-&LINER	\\
\enddata

\tablenotetext{a}{\tiny {\textbf {M.T.- Morphological Type. AGN classification denotes those galaxies that are AGN according to the [N II] diagrams but not to the [S II] and/or [O I].}}}  
\tablenotetext{\dag}{\tiny {\textbf {L-S classification means that galaxies fall in the separation line for Seyfert and LINER according to [S II] and [O I] diagrams.}}}
\tablenotetext{*}{\tiny {\textbf {Type with weak broad component in permitted lines. Seyfert quantitative classification according to \citet{1992MNRAS.257..677W}.}}}

\end{deluxetable}

\clearpage

\begin{deluxetable}{ccccccc} \tabletypesize{\footnotesize}
\tablecolumns{7} \tablewidth{0pt} \tablecaption{Logarithm intensities
ratios and their errors for the Varela's sample. Labels like on Table
6.}  \tablehead{ \colhead{Object}			 &
\colhead{M.T.\,\tablenotemark{a}}  	 &  \colhead{LOG([O
III]/H$\beta$)}      &  \colhead{LOG([N II]/H$\alpha$)} 	 &
\colhead{LOG([S II]/H$\alpha$)}      & \colhead{LOG([O I]/H$\alpha$) }
& \colhead{Type }	 } \startdata \sidehead{Elliptical} \hline
PGC29177&E-SO	&	-0.0385	$\pm$	0.0075	&	-0.4953	$\pm$
0.0064	&	-0.6389	$\pm$	0.0090	&	-1.9657	$\pm$
0.0284&HII	\\ PGC36037&E-SO	&	0.6757	$\pm$	0.2136
&	-0.0778	$\pm$	0.0767	&	-0.0145	$\pm$	0.1076	&
-0.8049	$\pm$	0.2415&S-L\dag	\\ PGC36211&E	&	0.1761	$\pm$
0.0310	&	-0.5803	$\pm$	0.0225	&	-0.2478	$\pm$	0.0225
&	-1.3349	$\pm$	0.0808&HII	\\ PGC43121&E-SO	&
0.4754	$\pm$	0.0048	&	-0.9475	$\pm$	0.0054	&
-0.8785	$\pm$	0.0066	&	-1.9157	$\pm$	0.0141&HII	\\

\sidehead{Spheroidal} \hline PGC25467&S0-a	&	-0.4046	$\pm$
0.0283	&	-0.3493	$\pm$	0.0119	&	-0.4011	$\pm$	0.0177
&	-1.5501	$\pm$	0.0721&HII	\\ PGC28259&S0	&	0.5808
$\pm$	0.0168	&	-0.4437	$\pm$	0.0169	&	-0.2733	$\pm$
0.0206	&	-0.8945	$\pm$	0.0330&Sy2	\\ PGC28758&S0-a
&	0.7331	$\pm$	0.0062	&	-1.2866	$\pm$	0.0097	&
-1.0391	$\pm$	0.0110	&	-2.0580	$\pm$	0.0283&HII	\\
PGC31601&S0	&	0.2531	$\pm$	0.0109	&	-0.5715	$\pm$
0.0092	&	-0.6821	$\pm$	0.0158	&	-1.4981	$\pm$
0.0367&HII	\\
\enddata 

\end{deluxetable}

\clearpage
\end{document}